# Membrane Tension Governs Particle Wrapping-Unwrapping Transitions and Stalling during Endocytosis

Yasin Ranjbar[1], Yujun Teng[2], Haleh Alimohamadi[3], Huajian Gao[4], Mattia Bacca[1,*]

[1]Mechanical Engineering, University of British Columbia, BC V6T 1Z4, Canada
[2]Mechanical and Aerospace Engineering, Nanyang Technological University, 639798, Singapore
[3]Molecular Biology and Biochemistry, University of California Irvine, 92697, USA
[4]Mechano-X Institute, Applied Mechanics Laboratory, Department of Engineering Mechanics, Tsinghua University, Beijing 100190, China

[*]Corresponding author: mbacca@mech.ubc.ca

**Abstract**

Membrane wrapping controls nanoparticle uptake during endocytosis, whereas the reverse process of membrane unwrapping accompanies particle expulsion and membrane fusion events. Existing theoretical descriptions typically focus on adhesion and bending energies within the particle–membrane contact region and often neglect the deformation energy of the membrane outside the contact zone. This approximation is valid only in the limit of vanishing membrane tension, where the non-contact membrane assumes a catenoid-like configuration with negligible bending energy. However, at finite tension the deformation of the non-contact membrane becomes a significant energetic contribution. Here we show that this tension-dependent non-contact energy governs the progression of particle wrapping. By analysing the variation of the total membrane energy with wrapping degree, we uncover a competition between particle adhesion, membrane tension and particle size that determines whether wrapping proceeds, stalls, or reverses into spontaneous unwrapping. This framework reveals a stalling boundary separating regimes of particle uptake and expulsion. To capture the non-contact deformation efficiently, we derive a compact phenomenological approximation that accurately reproduces the full numerical solution of the membrane shape. The resulting energetic map provides a unified physical description of particle wrapping and unwrapping, with implications for endocytosis, membrane fusion, and nanoparticle design.

## Introduction

Cells exchange nanoscale cargo with their surroundings through membrane-remodeling processes that can be broadly classified as wrapping and unwrapping (Fig. 1, top-right). These mechanisms govern the uptake and release of engineered nanoparticles, virions, and membrane-bound vesicles, and are therefore central to targeted drug delivery, viral infection, and vesicle-mediated cellular communication (Mitchell et al., 2021; Zhang et al., 2015; Helenius, 2018; Harrison, 2008). During wrapping, the cell membrane progressively encloses a particle, as in endocytosis or outward budding. During unwrapping, a membrane-coated cargo merges with a cellular membrane, as in viral fusion or vesicle exocytosis. Viral entry may occur through direct fusion with the plasma membrane or through endocytosis followed by fusion with an endosomal membrane (Mercer et al., 2010; Harrison, 2008). Vesicle fusion and exocytosis also underlie neurotransmitter release (Südhof 2004, Kaytanli 2025) and other forms of intercellular communication. In this work, 'wrapping' and 'unwrapping' are used as general mechanical descriptions of these membrane-mediated transport processes. Understanding the physical mechanisms governing wrapping and unwrapping is essential for predicting and controlling membrane-mediated transport across a broad range of biological and biomedical applications.

Over the past five decades, theoretical models in membrane biophysics and mechanics have been developed to describe these processes, which remain difficult to observe directly at the nanoscale. Most continuum models describe wrapping through a competition between adhesion energy, which drives membrane wrapping, and energetic penalties associated with membrane bending and tension (Helfrich, 1973; Seifert et al., 1991). If adhesion overcomes these penalties throughout the process, wrapping proceeds spontaneously; otherwise, it may stall at an intermediate state, favoring unwrapping, or fail to initiate. Membrane tension plays a particularly important role in regulating wrapping and unwrapping because, unlike adhesion and bending, it acts primarily through deformation of the membrane outside the particle–membrane contact region. Determining this contribution requires solving a nonlinear variational problem to obtain the membrane shape that minimizes the Helfrich free energy while satisfying the appropriate boundary conditions (Jülicher and Seifert, 1994; Deserno, 2004a). This procedure is computationally demanding, numerically sensitive, and unsuitable for rapid evaluation across broad parameter spaces (Deserno, 2004b). Consequently, many theoretical studies have neglected the free-membrane contribution or replaced it with simplified analytical approximations. Neglecting the free-membrane contribution is only justified for zero-tension membranes or near the initiation and completion of wrapping, where its energetic contribution becomes negligible. Earlier models represented it through an effective line tension at the contact line or explicit functions of the wrapping angle (Lipowsky and Döbereiner, 1998; Tzlil et al., 2004; Zhang et al., 2015; Bacca, 2022; Agostinelli et al., 2022), whereas Deserno (2004a) derived the exact zero-tension solution together with small-gradient and the high-tension asymptotic limit. Later, Foret (2014) and Frey et al. (2019) proposed increasingly accurate approximations for the loose- and tense-membrane regimes. Although these contributions have substantially advanced the understanding of membrane wrapping, they remain restricted to specific asymptotic regimes. At physiological membrane tensions, however, the free membrane contributes significantly to the overall energetics and strongly influences wrapping progression, highlighting the need for a simple, physically transparent, and uniformly accurate description across the full physiological range of membrane tensions.

In the present work, we show that membrane tension plays a fundamentally dual role during membrane-mediated transport. As illustrated schematically in Figure 1, wrapping undergoes a geometric transition from a peel-off regime before mid-wrapping to a seal-in regime thereafter. Before the midpoint, at $\theta < \pi/2$, further wrapping increases the extent of the non-flat membrane outside the particle, increasing the energetic penalty associated with membrane tension and bending and opposing further wrapping. Beyond the midpoint, for $\theta > \pi/2$, the opposite occurs: further wrapping progressively reduces the non-flat membrane region, causing membrane tension to promote wrapping instead. Thus, wrapping is limited by tension at first, and then it is promoted. The corresponding reversal process, unwrapping, is also subjected to the same initial penalty and subsequent aid by tension, leading to a unified physical picture for both membrane uptake and release.

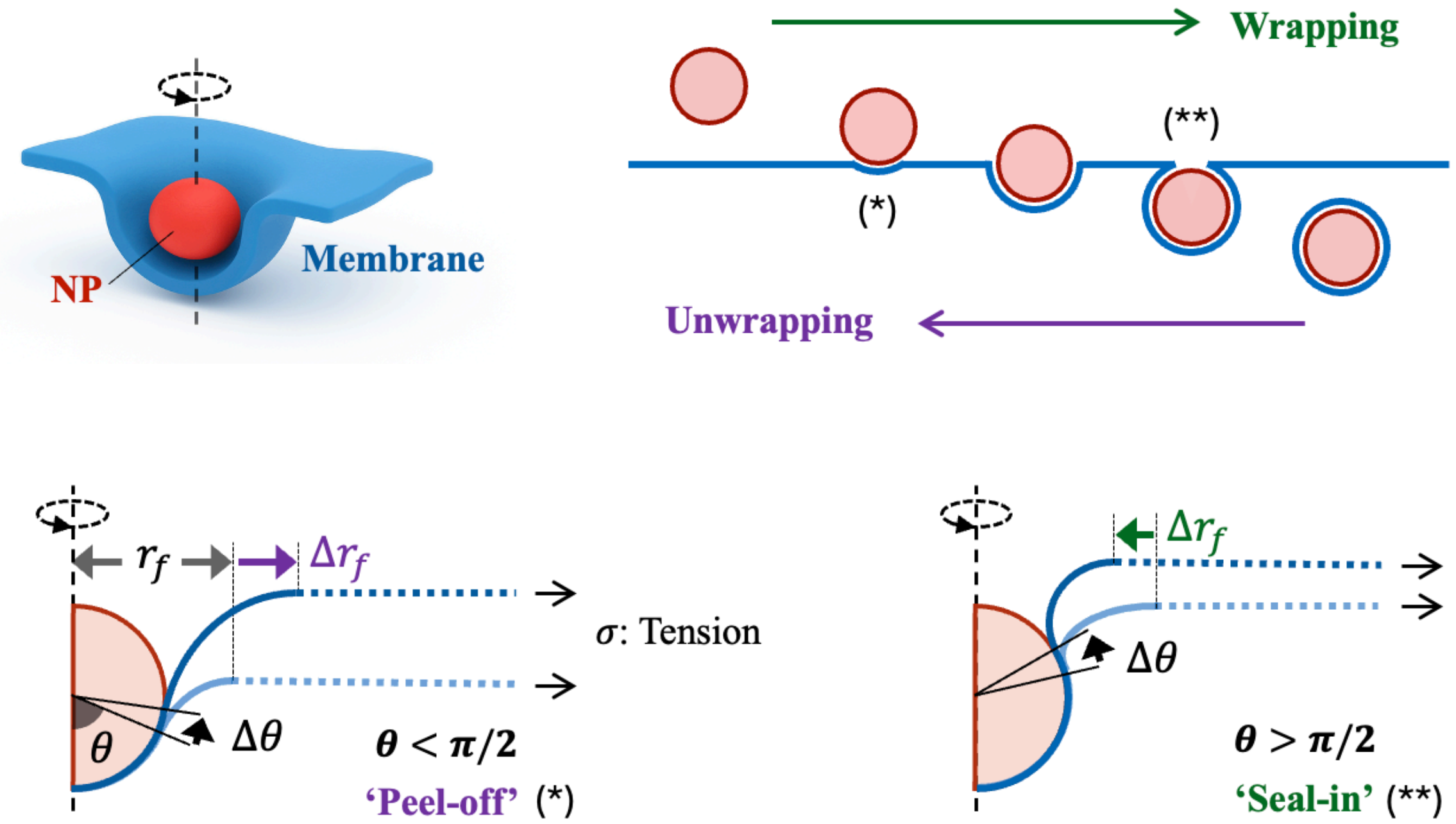

**Figure 1**: Schematic of nanoparticle (NP) wrapping and unwrapping by a membrane under tension. The top panels show endocytosis-like wrapping (green arrow), from left to right, where the (red) NP moves downward while becoming progressively engulfed by the (blue) membrane to form a vesicle-like endosome; and fusion-like unwrapping (purple arrow), from right to left, where a vesicle fuses its outer (blue) membrane with the cell membrane while the NP moves upward. The bottom panels illustrate two geometric regimes governing the evolution of the free membrane. In the ***'peel-off'*** regime, for $\theta < \pi/2$, an increment in wrapping advances the radial position $r_f$ of the connection between the deformed free membrane and the undeformed flat membrane (dashed line) outward. This increases the extent of the non-flat free membrane, raising the total free energy of the system and thereby hindering wrapping while promoting the reverse process, unwrapping. In the ***'seal-in'*** regime, for $\theta > \pi/2$, further wrapping instead drives $r_f$ inward. The consequent reduction of the non-flat free membrane lowers the total free energy of the system, thereby promoting wrapping and hindering unwrapping. These two geometric regimes underlie the dual energetic role of membrane tension, whereby both wrapping and unwrapping are opposed during their early stages but promoted toward completion.

Building on this physical picture, we derive a simple yet highly accurate closed-form expression for the free-membrane energy as a function of the wrapping angle and the dimensionless membrane tension, eliminating the need to repeatedly solve the membrane shape equations. The proposed formulation accurately reproduces the full numerical solution across the physiological range of membrane tensions while providing a transparent interpretation of the geometric origin of membrane-tension effects. By properly accounting for the energetics of the membrane outside the particle–membrane contact region, the framework predicts wrapping and unwrapping energy barriers and identifies conditions leading to stalled or incomplete wrapping. Furthermore, by evaluating the rate of change of the total energy with respect to wrapping progression, we determine the minimum adhesion energy required at every stage of wrapping to sustain spontaneous engulfment. This energetic criterion defines adhesion thresholds for complete wrapping, reveals when wrapping or unwrapping become energetically favorable, and provides a practical framework for analyzing and designing membrane-mediated transport processes.

### Energetics and Equilibrium Configuration of the Nanoparticle–Membrane System

Fig. 2 depicts the wrapping/unwrapping process of a rigid spherical nanoparticle (NP, red) by an initially flat, effectively infinite membrane (blue). The axisymmetric problem is described using Helfrich membrane theory (Helfrich 1973)

$$W = \int_0^{\infty} \left[\frac{B}{2}(H - H_0)^2 + B'K - \Gamma\right] 2\pi r ds + \sigma\Delta S + p\Delta V \tag{1}$$

Here, $B$ is the bending rigidity, penalizing deviations of the total curvature $H = C_1 + C_2$ from the spontaneous curvature $H_0$. The principal curvatures are $C_1$ and $C_2$, where $C_1$ is along the length of the membrane's curved profile in the radial plane, and $C_2$ is around the circumference in the rotational direction. $B'$ is the saddle-splay (or Gaussian) modulus, and $K = C_1 C_2$ is the Gaussian curvature. The free energy also includes the adhesion surface energy $\Gamma$ $[N/m]$ between the NP and membrane; a surface tension penalty $\sigma\Delta S$, where $\sigma$ $[N/m]$ is membrane tension, and $\Delta S = S - S_0$ is the increase in membrane area relative to its reference undeformed configuration; and a pressure penalty $p\Delta V$, where $p$ $[N/m^2]$ is the pressure difference across the membrane, and $\Delta V = V - V_0$ is the volume change relative to the reference state.

The equilibrium configuration of the membrane is described using an arc-length parameterization $s$, with $r(s)$ the distance from the *z*-axis of symmetry, $z(s)$ the height from the flat plane touching the bottom tip of the NP, and $\tan[\varphi(s)] = dz/dr$ the slope of the membrane with the horizontal plane (Fig. 2), such that

$$r_{,s} = \cos\varphi, \ \ z_{,s} = \sin\varphi \tag{2}$$

with $r_{,s} = dr/ds$, etc. The integration domain of Eq. (1) goes from $s = 0$ to $s = \infty$, encompassing three main regions of the membrane (see Fig. 2): the C domain from $s = 0$ to $s = s_i$; the NC domain from $s = s_i$ to $s = s_f$ and $\Gamma = 0$; and the flat (F) domain from $s = s_f$ to $s = \infty$, where also $\Gamma = 0$. Although the latter two domains can effectively be combined into one, we take $s_f$ as a finite length to avoid an infinite integration domain, given that a flat membrane has zero free energy. The principal curvatures in this axisymmetric geometry are (Seifert et al. 1991, Deserno 2004a, Yi *et al.* 2011, Napoli and Goriely, 2020)

$$C_1 = \varphi_{,s}, \ \ C_2 = \frac{\sin\varphi}{r} \tag{3}$$

For simplicity we assume zero spontaneous curvature throughout the membrane, $H_0 = 0$. The effect of spontaneous curvature, induced by asymmetries between inner and outer leaflets of the membrane or by protein domains, can readily be incorporated into the contact-region energy but is beyond the scope of the present study.

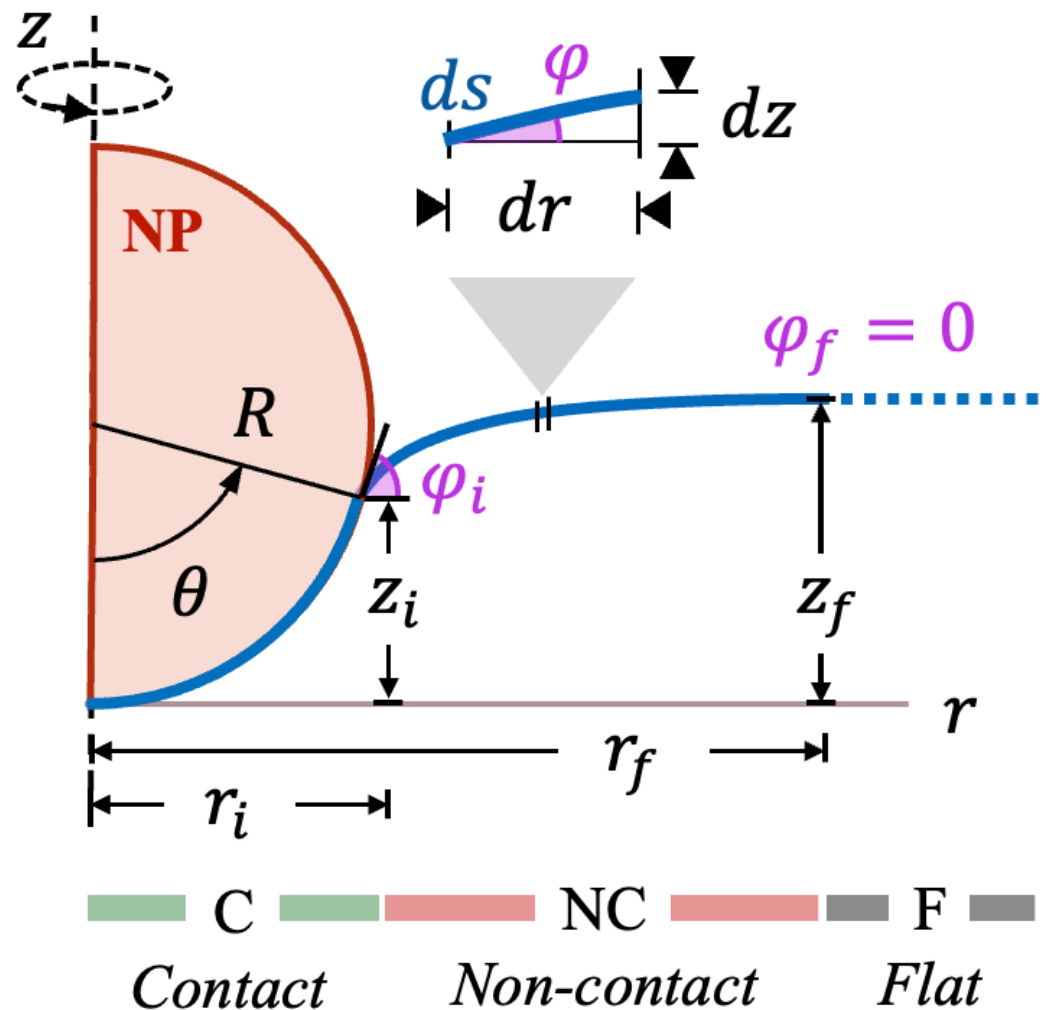

**Figure 2**: Axisymmetric geometry and notation of the membrane–nanoparticle system. The membrane is partitioned into three regions: the **contact (C)** region adhering to the nanoparticle, the **non-contact (NC)** region, corresponding to the non-flat free membrane between the contact line and the transition to the flat membrane, and the **flat (F)** region (dashed line), where the membrane is horizontal $(\varphi_f = 0)$. The contact line is located at $(r_i, z_i)$, while the transition between the NC and F regions occurs at $(r_f, z_f)$. The membrane profile is parameterized by the arc length $s$ and slope angle $\varphi$, with differential geometric relations $dr = ds\cos\varphi$ and $dz = ds\sin\varphi$. The nanoparticle has radius $R$, adopted as the characteristic length scale, and the wrapping state is characterized by the wrapping angle $\theta$, with $\theta = 0$ at wrapping initiation and $\theta = \pi$ at complete wrapping.

Using the Gauss-Bonnet theorem for axisymmetric surfaces, the integral of the Gaussian curvature in Eq. (1) over the entire membrane profile can be evaluated using contributions at the two ends of the axisymmetric membrane ($s = 0$ and $s = \infty$), which simplifies to

$$\int_0^\infty (B'K)\, 2\pi r\, ds = 2\pi B' \int_{\varphi(0)}^{\varphi(\infty)} \sin\varphi\, d\varphi \;=\; 0 \tag{4}$$

where the final equality holds due to the imposed BCs of $\varphi(0) = \varphi(\infty) = 0$ (discussed further below). Thus, the saddle-splay contribution always vanishes, and we can take $B' = 0$ everywhere (Seifert 1997, Deserno 2004a).

In our study, we take a nonzero membrane tension $\sigma$ everywhere, while we assume $p \approx 0$ for simplicity, corresponding to a perfectly permeable membrane. This assumption is common in the literature, and Yi et al. (2011) demonstrated that the volumetric energy term contributes only marginally to the equilibrium membrane configuration for pressure differences up to $p = 100$ Pa.

The membrane surface area is given by $S = \int_0^\infty 2\pi r ds$, while the reference (undeformed) surface is $S_0 = \int_0^\infty 2\pi r dr$. Using Eq. (2), the surface area change can be written as

$$\Delta S = \int_0^\infty 2\pi r(1 - \cos\varphi) ds \tag{5}$$

At this point, the free energy in Eq. (1) can be updated by plugging in Eq. (3) and (5), as well as setting $H_0 = 0$ and $p = 0$. The updated dimensionless form of the energy then becomes

$$\frac{W}{2\pi B} = \int_0^\infty \left[\frac{1}{2}\left(\varphi_{,\bar{s}} + \frac{\sin\varphi}{\bar{r}}\right)^2 - \bar{\Gamma} + \bar{\sigma}(1-\cos\varphi)\right]\bar{r}\, d\bar{s} \tag{6a}$$

where

$$\bar{s} = \frac{s}{R},\ \bar{r} = \frac{r}{R},\ \bar{z} = \frac{z}{R}, \bar{\Gamma} = \frac{\Gamma R^2}{B}, \bar{\sigma} = \frac{\sigma R^2}{B}, \varphi_{,\bar{s}} = \frac{d\varphi}{d\bar{s}} \tag{6b}$$

are dimensionless variables of the problem, with $R$ being the radius of the NP (Fig. 2).

To find the equilibrium configuration of the free membrane outside the C-domain (NC and F domains), we need to minimize its free energy as given by Eq. (6), while enforcing the arc-length parametrization in Eq. (2), which in dimensionless form reads as $\bar{r}_{,\bar{s}} = \cos\varphi$ and $\bar{z}_{,\bar{s}} = \sin\varphi$. We begin by expressing the energy outside the C region as

$$\frac{W_{NC}}{2\pi B} = \int_{\bar{s}_i}^\infty \mathcal{L}\, d\bar{s} \tag{7}$$

with $\mathcal{L}$ the dimensionless Lagrange energy density per unit length (Lagrangian) given by

$$\mathcal{L} = \left[\frac{1}{2}\left(\varphi_{,\bar{s}} + \frac{\sin\varphi}{\bar{r}}\right)^2 + \bar{\sigma}(1-\cos\varphi)\right]\bar{r} + \lambda_{\bar{r}}\left(\bar{r}_{,\bar{s}} - \cos\varphi\right) + \lambda_{\bar{z}}\left(\bar{z}_{,\bar{s}} - \sin\varphi\right) \tag{8}$$

Here, $\lambda_{\bar{r}}$ and $\lambda_{\bar{z}}$ are Lagrange multipliers that enforce the arc-length constraints in Eq. (2).

The equilibrium shape equations follow from the Euler-Lagrange (EL) equations (Seifert et al. 1991), which form a system of second-order ordinary differential equations (ODEs). While the EL equations correctly minimize $W_{NC}$, their second-order form is less convenient for numerical integration. For numerical solution it is convenient to reformulate the Euler–Lagrange equations in Hamiltonian form, yielding an equivalent first-order system.

The dimensionless Hamiltonian density per unit length (Hamiltonian) is defined in standard form as

$$\mathcal{H} = -\mathcal{L} + \sum q_{,\bar{s}}\, p_q$$

with $p_q = \partial\mathcal{L}/\partial q_{,\bar{s}}$ the conjugate momenta of the *coordinates* $q = \bar{r}$, $\bar{z}$, and $\varphi$, given by

$$p_{\bar{r}} = \lambda_r,\ p_{\bar{z}} = \lambda_z, \text{ and } p_\varphi = \bar{r}\varphi_{,\bar{s}} + \sin\varphi \tag{9}$$

yielding

$$\mathcal{H} = \frac{p_\varphi^2}{2\bar{r}} - \frac{p_\varphi \sin\varphi}{\bar{r}} + \bar{\sigma}(1-\cos\varphi)\,\bar{r} + p_{\bar{r}}\cos\varphi + p_{\bar{z}}\sin\varphi \tag{10}$$

Hamilton's equations are then

$$q_{,\bar{s}} = \frac{\partial\mathcal{H}}{\partial p_q},\ p_{q\,,\bar{s}} = -\frac{\partial\mathcal{H}}{\partial q} \tag{11}$$

Using Eq. (10) and (11), this yields the first-order system of ODEs

$$\bar{r}_{,\bar{s}} = \cos\varphi \tag{12a}$$

$$\bar{z}_{,\bar{s}} = \sin\varphi \tag{12b}$$

$$\varphi_{,\bar{s}} = \frac{p_\varphi}{\bar{r}} - \frac{\sin\varphi}{\bar{r}} \tag{12c}$$

$$(p_{\bar{r}})_{,\bar{s}} = \frac{p_\varphi^2}{2\bar{r}^2} - \frac{p_\varphi \sin\varphi}{\bar{r}^2} + \bar{\sigma}(1-\cos\varphi) \tag{12d}$$

$$(p_{\bar{z}})_{,\bar{s}} = 0 \tag{12e}$$

$$(p_\varphi)_{,\bar{s}} = \frac{p_\varphi \cos\varphi}{\bar{r}} + \bar{\sigma}\bar{r}\sin\varphi + p_{\bar{r}}\sin\varphi - p_{\bar{z}}\cos\varphi \tag{12f}$$

where Eq. (12a) and (12b) recover the arc-length parametrization.

Let us now divide the free energy in Eq. (1) into its adhesive contribution $-W_A$ in the C-domain, with negative sign indicating a release of free energy, driving the wrapping process; the contribution related to membrane deflection in the C-domain $W_C$, and that in the NC domain $W_{NC}$ (Fig. 2), giving the total as

$$W = -W_A + W_C + W_{NC} \tag{13}$$

Here we assume that the energy in the flat domain is zero. To calculate these energy contributions, we need to integrate Eq. (12), providing a closed form solution only for $W_A$ and $W_C$ in the C-domain, while the solution for $W_{NC}$ will be calculated numerically, and then fitted to an empirical function.

To ensure the spontaneity of the wrapping process, we must enforce a steady reduction of $W$ as wrapping progresses, *i.e.*,

$$\dot{W} = W_{,\theta}\dot{\theta} < 0 \tag{14a}$$

giving a release of free energy for a unit increment of the attachment angle $\theta$ (Fig. 1). This, from Eq. (13) implies

$$W_{A,\theta} > W_{C,\theta} + W_{NC,\theta} \tag{14b}$$

***Contact domain***: In the C-domain, the membrane is constrained to be attached to the rigid NP, and assumes the spherical cap configuration given by

$$\varphi = \bar{s}, \qquad \bar{r} = \sin\varphi, \qquad \bar{z} = 1 - \cos\varphi, \tag{15}$$

Thus, in this domain the membrane is not free to reach its equilibrium configuration, and so Eq. (12) does not apply here. This is because Eq. (12) does not account for the reaction forces exerted by the rigid NP on the membrane.

The energy in the C-region, $W_C$, is found by substituting Eq. (15) into (6), with integration bounds $\bar{s}$ between 0 and $\theta$, where $d\bar{s} = d\varphi$. $\theta$ is the attachment (or wrap) angle of the membrane with the NP (Fig. 2), ranging from 0 (no wrapping) to $\pi$ (completely wrapped) as the NP is endocytosed. The adhesion energy and membrane deflection energy are found to be

$$\frac{W_A}{2\pi B} = \bar{\Gamma}(1-\cos\theta) \tag{16a}$$

$$\frac{W_C}{2\pi B} = \left[2 + \frac{\bar{\sigma}}{2}(1-\cos\theta)\right](1-\cos\theta) \tag{16b}$$

With their energy rates

$$\frac{W_{A,\theta}}{2\pi B} = \bar{\Gamma}\sin\theta \tag{17a}$$

$$\frac{W_{C,\theta}}{2\pi B} = [2 + \bar{\sigma}(1 - \cos\theta)]\sin\theta \tag{17b}$$

***Non-Contact and Flat domains***: Here, the known BCs from the geometry (Fig. 2) are

$$\varphi_i = \theta, \qquad \bar{r}_i = \sin\theta, \qquad \bar{z}_i = 1 - \cos\theta \tag{18}$$

at the initial attachment with the NP, for $\bar{s} = \bar{s}_i = \theta$. At the F-region, for $\bar{s} \to \infty$, we impose the condition of asymptotic flatness (Deserno 2004a), such that

$$\lim_{\bar{s}\to\infty} \varphi(\bar{s}) = \lim_{\bar{s}\to\infty} \varphi_{,\bar{s}}(\bar{s}) = 0 \tag{19}$$

where the membrane becomes flat, approaching an undeformed state. To ensure the integration domain of Eq. (12) is finite, we replace the condition above with

$$|\varphi_f| \leq e_t \tag{20a}$$

$$\left|\left(\varphi_{,\bar{s}}\right)_f\right| \leq e_t \tag{20b}$$

at the far end of the NC region, for $\bar{s} = \bar{s}_f$. Here, $e_t$ is an error tolerance that is sufficiently small (discussed further below). Once Eq. (20) is satisfied, we assume that the membrane is now essentially "flat".

At the far end of the NC-domain ($\bar{s} = \bar{s}_f \to \infty$), $\bar{r}(\bar{s}_f)$ and $\bar{z}(\bar{s}_f)$ are unconstrained. Since the total NC energy ($W_{NC}$) is minimized for every membrane configuration through Eq. (12), the first-order change in energy with respect to arbitrary displacements in $\bar{r}$ and $\bar{z}$ must be zero. This stationarity condition yields two natural BCs at $\bar{s}_f$ (Jülicher and Seifert 1994):

$$p_{\bar{r}f} = 0 \tag{21a}$$

$$p_{\bar{z}f} = 0 \tag{21b}$$

Using Eq. (21b) and (12e), we can deduce that $p_{\bar{z}}(\bar{s}) = \lambda_{\bar{z}}(\bar{s}) = 0$ for all $\bar{s}$. As a result, Eq. (12e) is no longer needed, and both Eqs. (10) and (12f) must be updated to remove the $p_{\bar{z}}$ terms. Furthermore, the Hamiltonian density, $\mathcal{H}$ is conserved along the entire membrane, a condition that arises thanks to the Lagrangian $\mathcal{L}$ being explicitly independent of the arc-length $\bar{s}$ (Jülicher and Seifert, 1994). Further, by calculating $\mathcal{H}$ at the flat membrane (or equivalently at $\bar{s}_f$), from Eq. (10), (19), and (21), we get $\mathcal{H}_f = 0$, which we can extend to every point of the membrane, so $\mathcal{H} = 0$ everywhere. Knowing this, we can use Eq. (10) to get an expression for $p_{\bar{r}}$:

$$p_{\bar{r}} = -\frac{p_\varphi^2}{2\bar{r}\cos\varphi} + \frac{p_\varphi}{\bar{r}}\tan\varphi + \frac{\bar{\sigma}(1-\cos\varphi)\bar{r}}{\cos\varphi} \tag{22}$$

In this case, one might be tempted to substitute Eq. (22) into (12f), removing the need for Eq. (12d). However, we find that this substitution reduces the stability of our shape equations. As such, we instead use Eq. (22) to derive an expression for $p_{\bar{r}}$ at $\bar{s}_i$ only:

$$(p_{\bar{r}})_i = -\frac{(p_\varphi^2)_i}{2\sin\theta\cos\theta} + \frac{(p_\varphi)_i}{\cos\theta} + \bar{\sigma}(1 - \cos\theta)\tan\theta \tag{23}$$

With this, we are only missing the condition of $\left(p_\varphi\right)_i$, or equivalently $\left(\varphi_{,\bar{s}}\right)_i$ given the relation in Eq. (9), to begin the process of solving the system of ODEs in Eq. (12). The correct value of $\left(p_\varphi\right)_i$ must be determined through an implementation of the shooting method, where the system of ODEs is initialized using a guessed value of $\left(p_\varphi\right)_i$, and the equations are numerically integrated with the aim of achieving the asymptotic flatness conditions imposed in Eq. (20). The value of $\left(p_\varphi\right)_i$ is adjusted until Eq. (20) is satisfied and the configuration is that which yields the minimum $W_{NC}$ as well. The latter condition is technically not required but is an additional measure we take given the extreme sensitivity required to converge to the correct $\left(p_\varphi\right)_i$. Throughout our simulations, we find that even the slightest deviation ($< 0.1\%$) from the correct $\left(p_\varphi\right)_i$ can cause the membrane's configuration to diverge before achieving the asymptotic flatness condition (see also Figure 2 in Deserno 2004b). Our approach for dealing with the sensitivity of this problem is summarized below.

We first choose a trial range of $\left(p_\varphi\right)_i$ values, where for each $\left(p_\varphi\right)_i$ guess, we integrate the shape equations in Eq. (12) up to a large yet finite arc-length (e.g. $\bar{s} \approx 20$). During the integration, we calculate the dimensionless NC energy in Eq. (6) at each integration step, also checking for the condition of asymptotic flatness in Eq. (20). Generally, due to the delicacy of the solution, it is very common that none of the trials will satisfy Eq. (20). In that case, the search must be refined around the $\left(p_\varphi\right)_i$ value with the lowest total NC energy. This process is then repeated for progressively finer guesses of $\left(p_\varphi\right)_i$ until a solution that satisfies the asymptotic flatness condition is found. Additionally, at high wrapping angles ($\theta > \pi/2$) and large membrane tensions ($\bar{\sigma} \gtrapprox 5$), the numerical solutions may exhibit nonphysical features and reduced numerical reliability (see also Sec. III-B in Deserno, 2004a). Results in this regime are therefore reported but should be interpreted with caution.

To impose the condition in Eq. (20), we used $e_t = 10^{-3}$, as well as $10^{-6}$, obtaining nearly identical results for both the configurations and energies (<1% deviation), thus concluding $10^{-3}$ is a sufficiently small error tolerance. To solve the system of ODEs, we implemented an RK4 solver with a step size of $\Delta\bar{s} = 10^{-4}$. We also computed the configurations at $\Delta\bar{s} = 10^{-5}$, and found little to no change in our solutions, except an increase in computational time, hence, the former step size of $\Delta\bar{s} = 10^{-4}$ is sufficient. Additionally, in the initial guess range, it can be beneficial to integrate to multiple $\bar{s}_f$ values that approach our "large yet finite arc-length" (e.g. $\bar{s}_f = 5, 10, 15$ in addition to $\bar{s}_f = 20$), as in some instances, it is likely to get no valid energy values (*i.e.*, a diverging membrane) as we approach large integration lengths.

*Zero-Tension Case:* At zero membrane tension, the principal curvatures cancel out ($H = 0$), and the NC energy is zero. In this case, the shape of the NC membrane is given exactly by the catenoid solution of the axisymmetric shape equations (Deserno 2004a):

$$\bar{r}(\bar{z}) = r_0 \cosh\left(\frac{\bar{z}-z_0}{r_0}\right) \tag{24a}$$

where $r_0$ and $z_0$ are integration constants. These constants are uniquely determined by matching the radial coordinate $r$ and slope $dr/dz$ at the detachment point $\bar{s} = \bar{s}_i$. In our case of a spherical NP, we get

$$r_0 = \sin^2\theta \tag{24b}$$

$$z_0 = \bar{z}_i - r_0 \ln\left(\frac{1+\cos\theta}{\sin\theta}\right) \tag{24c}$$

## Results and Discussion

We first establish the physiologically relevant range of the dimensionless membrane tension, $\bar{\sigma} = \sigma R^2/B$. Experimental membrane tensions measured across different cell types by tether-pulling (Dai and Sheetz, 1999; Lieber et al., 2013) and optical techniques (Popescu et al., 2006) typically lie within the range $\sigma = 0 - 4.5 \times 10^{-4}\ N/m$ (Rangamani, 2022). A representative value of $\sigma = 2 \times 10^{-5}\ N/m$ is frequently adopted in membrane-wrapping studies (Deserno, 2004a). The membrane bending rigidity is typically taken as $B \approx 20\ k_B T$ ($8.56 \times 10^{-20}\ J$), consistent with Helfrich's original estimate and widely used theoretical studies (Helfrich, 1973; Deserno, 2004a; Gao et al., 2005), although experimentally reported values span approximately $1 - 35\ k_B T$, depending on membrane composition and measurement conditions.

The present model assumes an initially flat membrane, which is appropriate when the cell radius greatly exceeds the nanoparticle radius ($R_{\text{cm}} \gg R$). Giant cells typically have radii of several micrometers, whereas nanoparticles undergoing receptor-mediated endocytosis commonly have radii between 5 and 50 nm, with an experimentally observed optimum near 25 nm (Aoyama et al., 2003; Gao et al., 2005; Chithrani et al., 2006, 2007). Particles substantially larger than this range fall outside the regime where the flat-membrane approximation is expected to remain accurate. We therefore adopt 50 $nm$ as the upper representative particle size throughout the present study.

The particle-membrane adhesion energy is the remaining key control parameter. Experimental measurements report adhesion energies spanning five orders of magnitude, from approximately $\Gamma \approx 10^{-8}\ N/m$ for weakly adhering lipid bilayers to $\Gamma \approx 10^{-3}\ N/m$ for strongly adhesive interfaces (Gruhn et al., 2007; Schönherr et al., 2004; Anderson et al., 2009; Agudo-Canalejo and Lipowsky, 2015). Adhesion to citrate-capped gold nanoparticles may be even stronger because of the large Hamaker constant of gold (Liu, 2016).

**Table 1**: Representative physical parameters and ranges adopted in our analysis

| Parameter | Symbol | Range/Value | Units |
|---|---|---|---|
| *Bending Rigidity* | $B$ | $20\ (8.56 \cdot 10^{-20})$ | $k_B T\ (J)$ |
| *Membrane Tension* | $\sigma$ | $0 - 4.5 \cdot 10^{-4}$ | $N/m$ |
| *NP Radius* | $R$ | $5 - 50 \cdot 10^{-9}$ | $m$ |
| *Adhesion Surface Energy* | $\Gamma$ | $10^{-8} - 10^{-3}$ | $N/m$ |

The representative physical parameters adopted throughout this study are summarized in Table 1. These values yield a maximum dimensionless membrane tension of approximately $\bar{\sigma} \approx 13.1$ and a maximum dimensionless adhesion of $\bar{\Gamma} \approx 29.2$. Throughout this work, we therefore consider the range $0 \leq \bar{\sigma} \leq 10$, which spans physiologically relevant membrane tensions while remaining representative of experimentally accessible nanoparticle sizes and membrane properties. Taking $\sigma = 2 \times 10^{-5}$ N/m, $B = 20\ k_B T$, and $R = 25\ nm$ as representative values gives $\bar{\sigma} \approx 0.15$. As shown below, even at the upper limit of $\bar{\sigma} = 10$, the adhesion required for complete wrapping remains well within the experimentally accessible range.

We characterize the progression of wrapping using the normalized wrapping fraction,

$$f = \frac{\theta}{\pi} \tag{25}$$

where $\theta$ is the membrane attachment angle (Fig. 2). Thus, $f = 0$ corresponds to the onset of wrapping, whereas $f = 1$ denotes complete engulfment of the nanoparticle.

Fig. 3 shows the equilibrium membrane configurations obtained by solving Eq. (12) for two wrap angles, $\theta = 0.3\pi$ and $0.7\pi$, corresponding to wrapping fractions of 30% ($f = 0.3$) and 70% ($f = 0.7$), respectively. Profiles are shown for dimensionless membrane tensions $\bar{\sigma} = \sigma R^2/B = 0,\ 0.5,\ 2,$ and 10. In the tensionless limit ($\bar{\sigma} = 0$), the free membrane assumes the exact catenoid solution, characterized by zero total curvature, $H = C_1 + C_2 = 0$ (Eq. (3)), everywhere along the non-contact membrane (black dashed line). Consequently, the free membrane has zero bending energy, and the NC contribution to the total energy vanishes. Because the catenoid approaches a flat configuration only asymptotically, the zero-tension membrane never reaches a flat state at any finite distance from the nanoparticle. Although this idealization is recovered only for vanishing membrane tension, it provides a useful limiting case and can be reproduced exactly using the analytical solution of Eq. (24).

For all finite values of $\bar{\sigma}$, the membrane instead approaches a flat configuration over a finite length scale. The solid curves in Fig. 3 represent the numerically resolved membrane profiles within the NC domain, whereas the dashed horizontal segments denote the asymptotically flat membrane in the F domain. The matching point between these two regions is determined numerically using the asymptotic flatness criterion of Eq. (20), terminating the integration once both the membrane slope angle and its derivative satisfy $|\ \phi\ | \leq e_t$ and $|\ d\phi/d\bar{s}\ | \leq e_t$, with $e_t = 10^{-3}$. This matching location therefore depends on the prescribed numerical tolerance and should not be interpreted as a unique physical transition point; a smaller tolerance would simply extend the numerical solution farther into the far field. Nevertheless, because the same criterion is applied for every value of $\bar{\sigma}$, Fig. 3 provides a consistent qualitative comparison demonstrating that increasing membrane tension progressively localizes the membrane deformation, reducing the spatial extent of the non-contact membrane.

In the asymptotic limit $\bar{\sigma} \to \infty$, the deformation becomes confined to an increasingly narrow boundary layer adjacent to the nanoparticle, while the remainder of the membrane remains essentially flat. Equivalently, this corresponds to $B/(\sigma R^2) \to 0$, achieved for vanishing bending rigidity or indefinitely large membrane tension or particle size. Although this limiting case is not physically attainable, it provides a useful interpretation of the localization mechanism. The upper bound $\bar{\sigma} = 10$ adopted here is also physically reasonable, as it follows from the representative ranges of membrane tension, bending rigidity, and nanoparticle size discussed at the beginning of the Results and Discussion section. Thus, the range $\bar{\sigma} = 0 - 10$ spans the relevant parameter space while approaching the high-tension localization regime.

It should be noted that, although increasing membrane tension reduces the spatial extent of the NC domain (Fig. 3), the membrane free energy itself scales with tension. Consequently, the limit $B/(\sigma R^2) \to 0$, despite producing a vanishingly small NC domain, is still associated with a significant deformation energy. This indicates that the increase in average membrane free-energy density with tension outweighs the reduction in the extent of the deformed membrane. *Appendix* $A$ reports the NC free energy, $W_{NC}$, for the membrane configurations shown in Fig. 3 over the full wrapping range, $f = 0 - 1$ (equivalently, $\theta = 0 - \pi$). As shown in Fig. A1, $W_{NC}$ vanishes at $f = 0$ and $f = 1$, reaches a maximum at an intermediate wrapping fraction, and increases approximately in proportion to $\bar{\sigma}$. The numerical results are further represented by an empirical beta-distribution approximation, whose coefficients are expressed as functions of the dimensionless membrane tension. This compact representation accurately reproduces the numerical solutions over the explored parameter range and provides a convenient closed-form approximation for evaluating the NC energy in endocytosis models without repeatedly solving the membrane shape equations. For example, the numerical results of Zhang *et al.* (2009) over the range $\bar{\sigma} \approx 0.02 - 0.2$ are reproduced accurately by Eqs. (A1)–(A5).

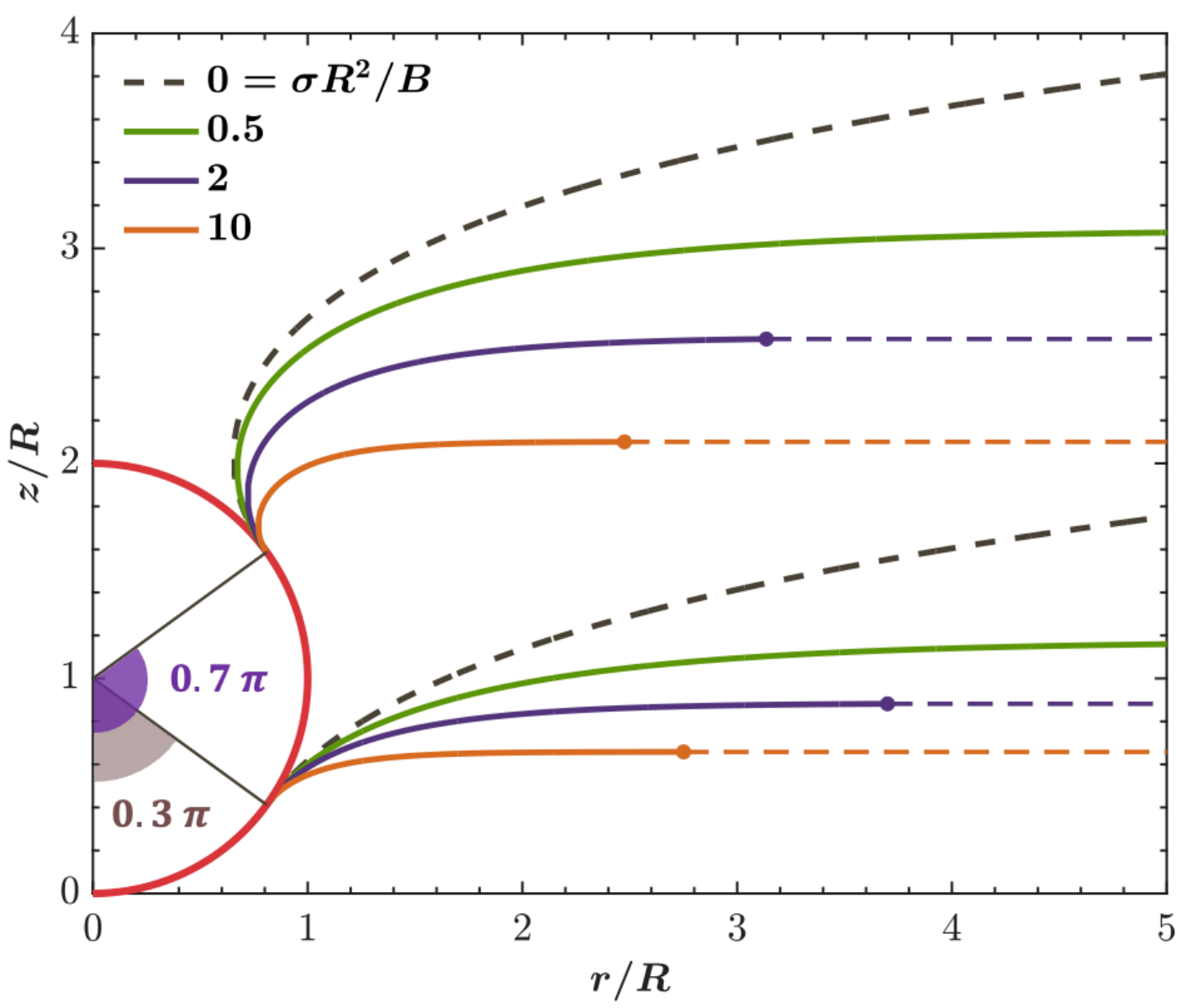

**Figure 3**: Equilibrium membrane profiles around a rigid spherical NP (red) for wrap angles $\theta = 0.3\pi$ and $0.7\pi$, shown for different values of $\sigma R^2/B$. Profiles are plotted in normalized coordinates $r/R$ and $z/R$. In the tensionless case ($\sigma = 0$), the membrane adopts a catenoid shape with zero NC energy and does not reach a flat configuration at any finite distance (black dashed curves). For finite $\sigma$, the membrane approaches a flat configuration over a finite length scale, with the transition becoming progressively sharper as $\sigma R^2/B$ increases. The solid curves denote the numerically resolved membrane profiles in the NC domain, whereas the dashed horizontal segments represent the asymptotically flat, zero-free-energy membrane in the F domain. The marked points indicate the numerical matching locations between the NC and F domains, defined by terminating the integration once the asymptotic flatness criterion of Eq. (20), $|\ \phi\ | \leq e_t$ and $|\ d\phi/d\bar{s}\ | \leq e_t$, is satisfied. Throughout this work, we adopt $e_t = 10^{-3}$, which yields membrane configurations and energies differing by less than 1% from those obtained with $e_t = 10^{-6}$. The matching locations therefore depend on the prescribed tolerance $e_t$ and should not be interpreted as unique physical transition points. Nevertheless, because the same convergence criterion and tolerance are applied for all values of $\sigma R^2/B$, the comparison consistently demonstrates that increasing membrane

tension localizes the deformed free membrane, thereby reducing the spatial extent of the NC domain.

In Fig. 4, we plot the total dimensionless membrane free energy, $\overline{W} = \overline{W}_C + \overline{W}_{NC}$, with $\overline{W} = W/(2\pi B)$, as a function of wrapping fraction $f$ (Eq. (25)) for $\bar{\sigma} = \sigma R^2/B = 0.125, 0.25, 0.5, 1, 2, 4, 5,$ and $10$. The markers represent the numerically computed NC energy, $\overline{W}_{NC}$, added to the closed-form C-domain contribution, $\overline{W}_C$, given by Eq. (16b). The solid curves combine $\overline{W}_C$ with the fitted approximation for $\overline{W}_{NC}$ developed in *Appendix A*, whereas the dashed curves show $\overline{W}_C$ alone, corresponding to neglecting the energy stored in the NC domain. Solid and dashed curves coincide at $f = 0$ and $f = 1$, where the NC contribution vanishes, but differ substantially at intermediate wrapping fractions. This discrepancy shows that neglecting the NC energy leads to a significant underestimation of the energetic cost of wrapping. Under the conditions considered here, the NC contribution accounts for up to approximately 40% of the total membrane free energy at intermediate wrapping fractions and therefore cannot be neglected when evaluating the energetic barrier to endocytosis.

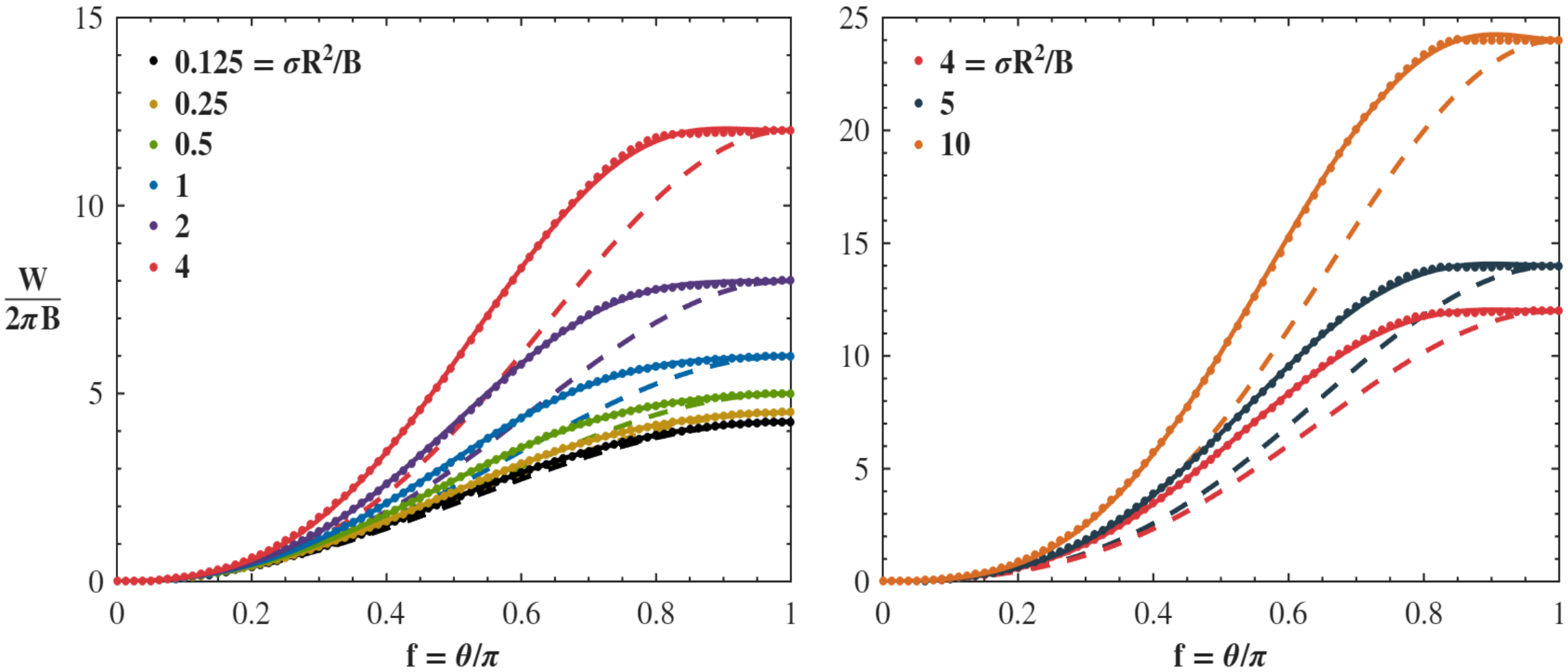

**Figure 4**: Total dimensionless membrane free energy, $W/(2\pi B)$, as a function of wrapping fraction $f$ (Eq. (25)) for different values of the dimensionless membrane tension $\bar{\sigma} = \sigma R^2/B$. Markers denote the total energy obtained by adding the numerically computed non-contact contribution, $W_{\mathrm{NC}}$, to the analytical contact-domain contribution, $W_{\mathrm{C}}$. Solid curves combine the analytical contact-domain energy with the fitted approximation for $W_{\mathrm{NC}}$ presented in *Appendix A*. Dashed curves show the contact-domain contribution alone and therefore neglect the energy stored in the non-contact membrane. The difference between the solid and dashed curves quantifies the contribution of the non-contact membrane to the total energy, which is most pronounced at intermediate wrapping fractions.

Spontaneous wrapping requires the inequality in Eq. (14) to be satisfied. Combining Eq. (14) with Eq. (16) gives

$$\bar{\Gamma} > \bar{\Gamma}^* \tag{26a}$$

where

$$\bar{\Gamma}^* = 2 + \bar{\sigma}(1 - \cos\theta) + \frac{1}{2\pi^2 B \sin\theta}\frac{\partial W_{NC}}{\partial f} \tag{26b}$$

is the minimum dimensionless adhesion required at each stage of wrapping. In deriving Eq. (26b), we used Eq. (25) and (17b).

While the C-domain contribution to Eq. (26b) increases monotonically with wrapping, the contribution arising from the NC membrane is non-monotonic, as reflected in Fig. 5. Together with the geometric factor $1/\sin\theta$, which accounts for the rate at which new contact area is created, this contribution causes the local adhesion requirement to increase and subsequently decrease, producing distinct maxima and minima. *Appendix A* reports the corresponding membrane-energy rate used to evaluate Eq. (26b).

Fig. 5 reports the minimum dimensionless adhesion required for spontaneous wrapping, $\bar{\Gamma}^* = \Gamma^* R^2/B$, as a function of the wrapping fraction $f$, for $\bar{\sigma} = 0.125, 0.5, 2,$ and 4. The adhesion requirement is computed from Eq. (26b) by substituting the numerical values of $\partial W_{NC}/\partial f$. Symbols denote the numerical results, whereas the dashed curves correspond to the classical approximation obtained by neglecting the NC-domain contribution ($W_{NC} \approx 0$). Unlike the latter, the present model predicts a pronounced maximum adhesion requirement, $\bar{\Gamma}^*_{\max}$, occurring at approximately $f = 0.5 - 0.6$ for all membrane tensions. This maximum originates entirely from the non-monotonic contribution of the NC membrane and therefore cannot be captured when $W_{NC}$ is neglected.

Assuming a uniform particle-membrane adhesion energy density, complete spontaneous wrapping requires $\bar{\Gamma} > \bar{\Gamma}^*_{\max}$, so that Eq. (14) is satisfied throughout the entire wrapping process. Conversely, when $\bar{\Gamma} < \bar{\Gamma}^*_{\max}$ only portions of the wrapping pathway satisfy the energetic condition for spontaneous uptake, whereas the remaining portions favor the reverse process. As a result, the system becomes trapped in a metastable partially wrapped configuration, giving rise to wrapping stalling.

The minimum adhesion required to initiate wrapping is $\bar{\Gamma} = 2$, corresponding to $\Gamma = 2B/R^2$, independent of membrane tension. This threshold equals the bending-energy density required to transform an initially flat membrane into a spherical surface of radius $R$. Because the membrane is initially flat, the contribution of membrane tension vanishes at the onset of wrapping, leaving bending as the sole energetic penalty. The same argument applies at complete wrapping, where the surrounding membrane again becomes nearly flat, explaining why the adhesion requirement returns to $\bar{\Gamma} = 2$ at $f = 1$.

To illustrate these predictions, Fig. 5 also shows two representative adhesion levels, $\bar{\Gamma} = 6$ and 0.5, for the case $\bar{\sigma} = 4$. Green segments identify regions where wrapping is spontaneous ($\bar{\Gamma} > \bar{\Gamma}^*$), whereas red segments indicate regions where the reverse process is energetically favored. The arrows indicate the direction of wrapping ($f: 0 \rightarrow 1$) and unwrapping ($f: 1 \rightarrow 0$).

For $\bar{\Gamma} = 6$, wrapping proceeds spontaneously up to approximately $f = 0.33$, where it stalls because the available adhesion becomes insufficient to overcome the membrane-deformation energy. Additional energy input, for example from clathrin recruitment or other active cellular processes, would be required to drive the system beyond this barrier. Once wrapping reaches approximately $f = 0.74$, spontaneous uptake resumes and proceeds autonomously to completion. The width of the stalled interval is controlled by the difference between the available adhesion and the peak requirement $\bar{\Gamma}^*_{\max}$, indicated by the upper star in Fig. 5. The numerical values of $\bar{\Gamma}^*_{\max}$ increase approximately linearly with membrane tension and are reported in Fig. 6-left, where they are fitted by

$$\bar{\Gamma}^*_{max} = 2 + 1.67\bar{\sigma}^{0.97} \tag{27}$$

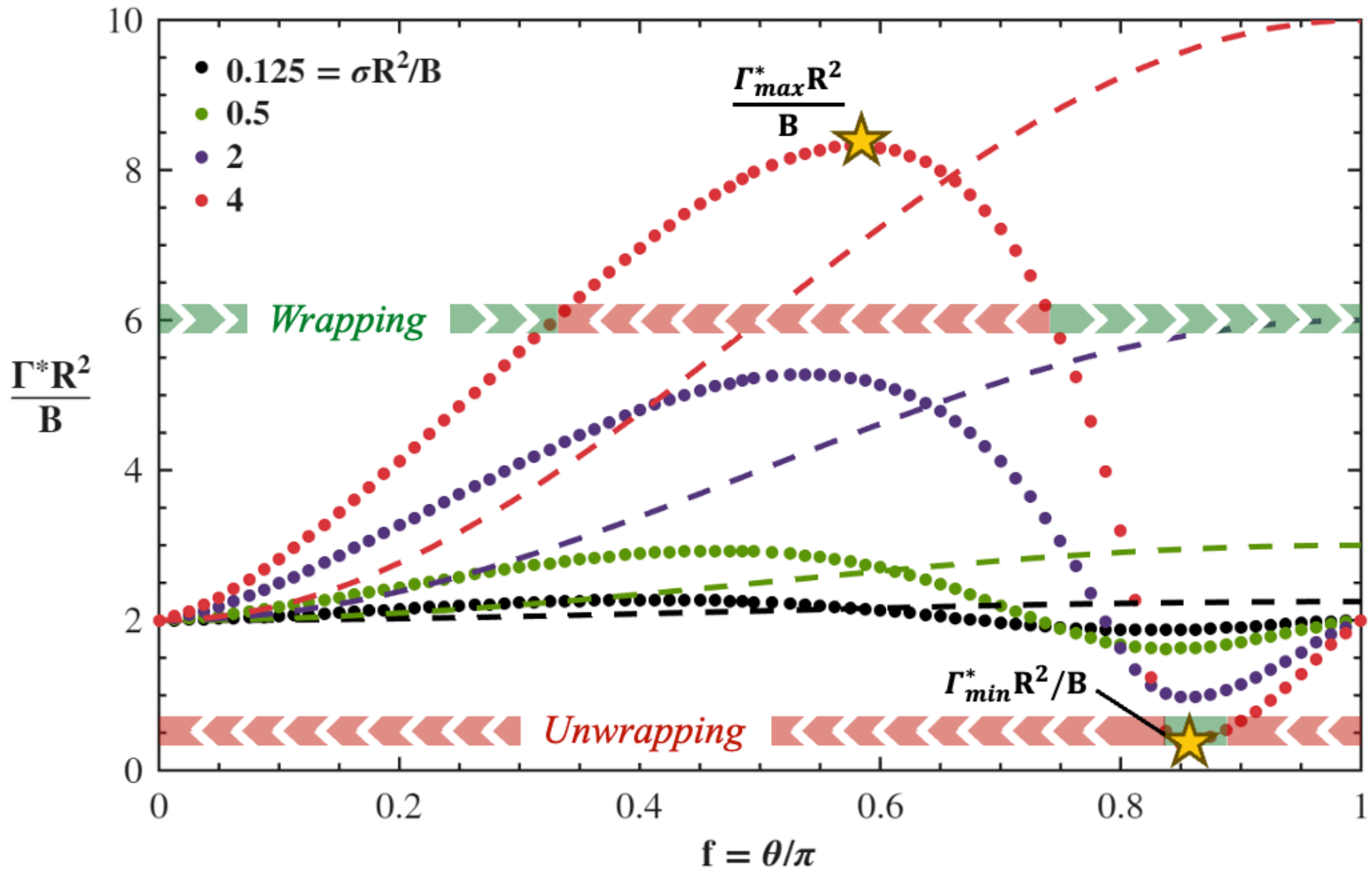

**Figure 5**: Dimensionless local adhesion requirement, $\bar{\Gamma}^* = \Gamma^* R^2/B$, as a function of wrapping fraction $f$ (Eq. (25)), computed from Eq. (26b) for different values of the dimensionless membrane tension $\bar{\sigma} = \sigma R^2/B$. Markers denote the numerical results, whereas dashed curves show the approximation obtained by neglecting the non-contact membrane contribution, $W_{\mathrm{NC}} \approx 0$. For the representative adhesion level $\bar{\Gamma} = 6$, wrapping is spontaneous where $\bar{\Gamma} > \bar{\Gamma}^*$ (green), whereas $\bar{\Gamma} < \bar{\Gamma}^*$ (red) indicates energetically unfavorable wrapping. The local maximum, $\bar{\Gamma}^*_{\max}$, defines the adhesion threshold required for uninterrupted wrapping, while the local minimum, $\bar{\Gamma}^*_{\min}$, defines the maximum adhesion compatible with uninterrupted unwrapping. The intervals in which the imposed adhesion lies between these extrema correspond to metastable partially wrapped states and wrapping or unwrapping stalling.

For the weaker adhesion level $\bar{\Gamma} = 0.5$, wrapping cannot initiate because the universal initiation threshold, $\bar{\Gamma} = 2$, is not satisfied. The energetically preferred direction is therefore unwrapping. Nevertheless, wrapping remains energetically favorable over approximately $f = 0.83 - 0.87$, causing spontaneous unwrapping to stall upon reaching this interval from $f = 1$. This behavior may provide a mechanical explanation for kiss-and-run fusion, in which transient fusion pores reseal before vesicles proceed to complete fusion, as observed experimentally in nerve terminals and neuronal synapses (Klyachko and Jackson, 2002; He et al., 2006). Unwrapping without stalling requires $\bar{\Gamma} < \bar{\Gamma}^*_{\min}$, where $\bar{\Gamma}^*_{\min}$ is the minimum adhesion requirement indicated by the lower star in Fig. 5. Its dependence on membrane tension is shown in Fig. 6-right and fitted by

$$\bar{\Gamma}^*_{min} = 2 - 0.62\bar{\sigma}^{0.69} \quad (28)$$

When $\bar{\Gamma}^*_{\min} < 0$, spontaneous unwrapping necessarily stalls for any positive adhesion energy. Completion of unwrapping would then require an effective negative adhesion, which could arise from repulsive interactions between the membrane and the nanoparticle or its contents.

The above findings complete the physical picture introduced in Fig. 1. For $\theta < \pi/2$, motion of the contact line occurs through a **peel-off** geometry, in which membrane tension promotes detachment of the membrane from the nanoparticle. Consequently, tension *hinders wrapping but promotes unwrapping*. Once $\theta > \pi/2$, the geometry changes to **seal-in**, where membrane tension promotes attachment of the membrane to the nanoparticle. In this regime, tension *promotes wrapping but hinders unwrapping*. Thus, membrane tension reverses its influence at the halfway point, providing a unified explanation for both the enhanced completion of wrapping beyond half engulfment and the stalling of reverse unwrapping, which may underlie kiss-and-run fusion.

Fig. 6 summarizes the practical design rules emerging from the present analysis by correlating the critical adhesion thresholds for uninterrupted wrapping and unwrapping with membrane tension. The non-contact membrane contribution affects the two processes asymmetrically because its energy varies non-monotonically with wrapping degree. During the early stages of wrapping, the NC deformation increases the local adhesion requirement to complete wrapping without stalling. At larger wrapping fractions, however, the NC energy decreases, reducing the local energetic cost of further wrapping while simultaneously creating an energetic barrier against reverse unwrapping. Consequently, the critical adhesion for uninterrupted wrapping, $\bar{\Gamma}^*_{\max}$, increases nearly linearly with membrane tension, whereas the critical adhesion below which spontaneous unwrapping can proceed, $\bar{\Gamma}^*_{\min}$, decreases. Thus, increasing membrane tension progressively strengthens the resistance to reverse unwrapping and enlarges the parameter regime in which partially wrapped states remain stable, providing a mechanical basis for kiss-and-run fusion. The fitted correlations given by Eqs. (27) and (28) accurately reproduce the numerical results over the investigated tension range and provide convenient closed-form expressions for estimating the onset of wrapping and unwrapping stalling without repeatedly solving the membrane shape equations. In contrast, neglecting the non-contact membrane contribution predicts a constant wrapping threshold and completely misses the tension-induced stabilization of partially wrapped states.

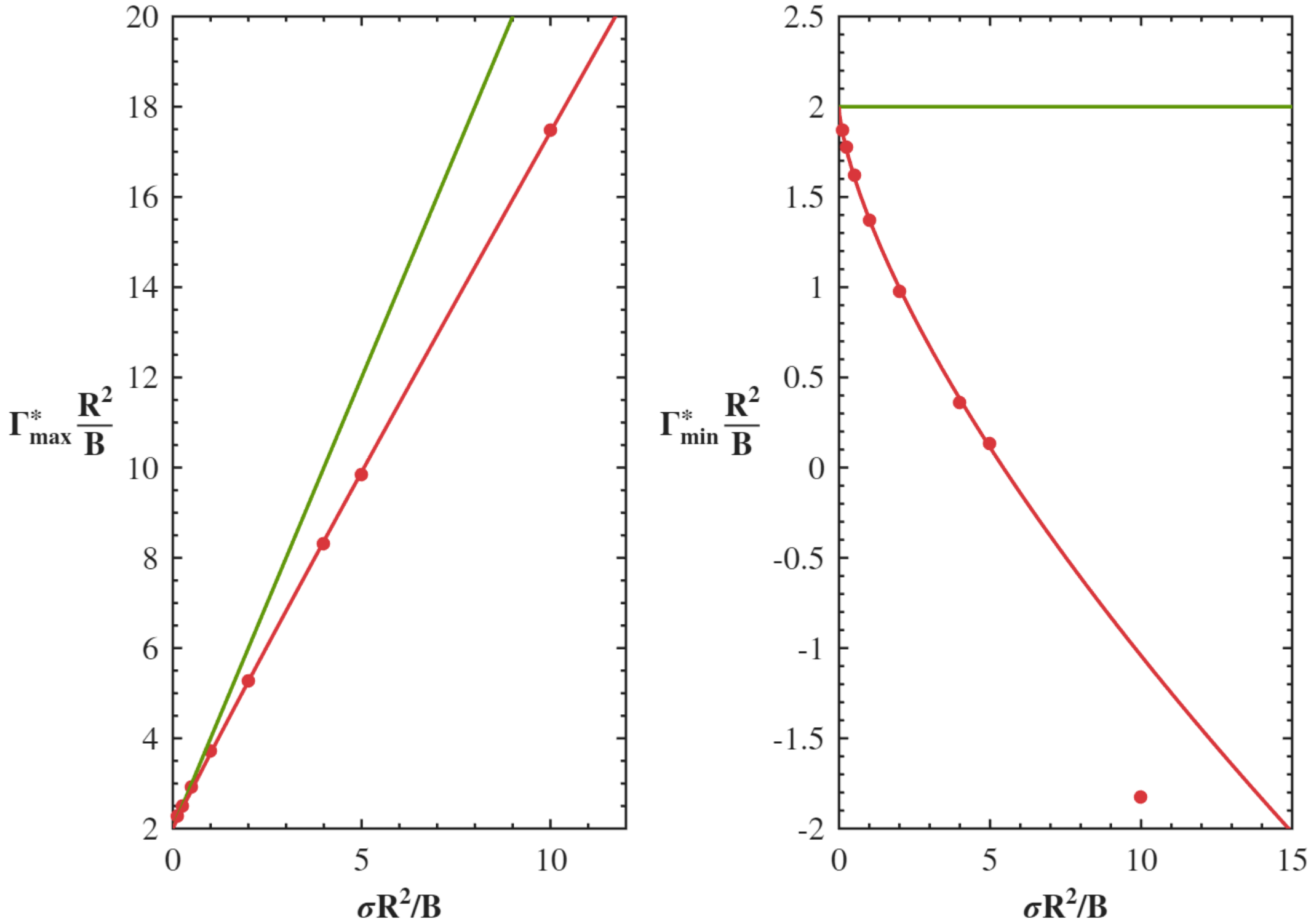

**Figure 6**: Critical adhesion thresholds for uninterrupted wrapping and unwrapping as functions of the dimensionless membrane tension, $\bar{\sigma} = \sigma R^2/B$. ***Left*****:** Maximum local adhesion requirement, $\bar{\Gamma}^*_{\max} = \Gamma^*_{\max} R^2/B$, above which wrapping proceeds without stalling. ***Right***: Minimum local adhesion requirement, $\bar{\Gamma}^*_{\min} = \Gamma^*_{\min} R^2/B$, below which unwrapping proceeds without stalling. Red markers denote numerical results, and red solid lines show the fitted correlations given by Eqs. (27) and (28). Green lines show the corresponding predictions obtained by neglecting the non-contact membrane contribution.

## Conclusion

In this work, we investigated the role of the non-contact (NC) membrane during the wrapping and unwrapping of a rigid spherical nanoparticle under membrane tension. By solving the full axisymmetric membrane shape equations and mapping the variation of the total membrane free energy with wrapping degree, we showed that the deformation of the NC membrane constitutes a major and intrinsically non-monotonic energetic contribution over the physiological range of membrane tensions. Although increasing membrane tension progressively localizes the NC deformation to a narrow region adjacent to the contact line, the associated deformation energy remains significant and therefore cannot be neglected when describing membrane-mediated transport.

The proposed framework reveals that membrane tension governs wrapping not only through the total deformation energy but, more importantly, through the local adhesion requirement needed to advance the membrane contact line. This requirement is determined by the rate of membrane-energy variation together with the rate at which new particle–membrane contact area is created. As a consequence, membrane tension plays a fundamentally dual role: it initially opposes wrapping while promoting unwrapping through a peel-off mechanism, but beyond approximately half wrapping it instead promotes wrapping and hinders unwrapping through a seal-in mechanism. This geometric transition provides a unified physical explanation for the reversal of tension effects during membrane uptake and release.

These competing mechanisms generate local maxima and minima in the adhesion requirement, giving rise to wrapping and unwrapping stalling, metastable partially wrapped configurations, and hysteresis between uptake and release pathways. In particular, the predicted resistance to reverse unwrapping at large wrapping fractions offers a mechanical explanation for the experimentally observed kiss-and-run mode of vesicle fusion, in which transient fusion pores reseal before complete fusion occurs.

Finally, we developed a compact closed-form approximation for the NC membrane energy that accurately reproduces the numerical solutions over the investigated parameter range, together with simple correlations for the critical adhesion thresholds governing uninterrupted wrapping and unwrapping. These expressions eliminate the need to repeatedly solve the membrane shape equations while providing practical design rules for predicting membrane-mediated particle uptake, expulsion, and fusion. While previous attempts provided simplified formulations for the influence of membrane tension on the total free energy, their accuracy was generally restricted to asymptotic regimes of very low or very high tension. In contrast, the present approximation remains accurate at intermediate tensions, which are representative of many biologically relevant conditions. More broadly, the present framework establishes a unified energetic description of wrapping and unwrapping that may facilitate the analysis and engineering of endocytosis, membrane fusion, and nanoparticle-based delivery systems.

## Appendix A: Membrane Energy in the NC Domain

The free energy of the membrane in the NC-domain can be approximated by the function

$$W_{NC} = W_{NC}^{*}\, g(f) \tag{A1}$$

where $W_{NC}^{*}$ is the maximum energy achieved at the critical wrapping degree $f^{*}$, and

$$g(f) = \left(\frac{f}{f^{*}}\right)^{kf^{*}} \left(\frac{1-f}{1-f^{*}}\right)^{k(1-f^{*})} \tag{A2}$$

is a dimensionless function that takes the form of a beta distribution, with $k$ the shape factor of the distribution. $W_{NC}^{*}$, $f^{*}$ and $k$ depend on the dimensionless tension $\bar{\sigma}$, via

$$\frac{W_{NC}^{*}}{2\pi B} = a\bar{\sigma}^{\alpha} \tag{A3}$$

$$f^{*} = 1 - \frac{1}{2\left(1+b\bar{\sigma}^{\beta}\right)} \tag{A4}$$

$$k = c + \frac{d}{e+\bar{\sigma}} \tag{A5}$$

where $a$, $b$, $c$, $d$, $e$, $\alpha$, and $\beta$ are fitting constants.

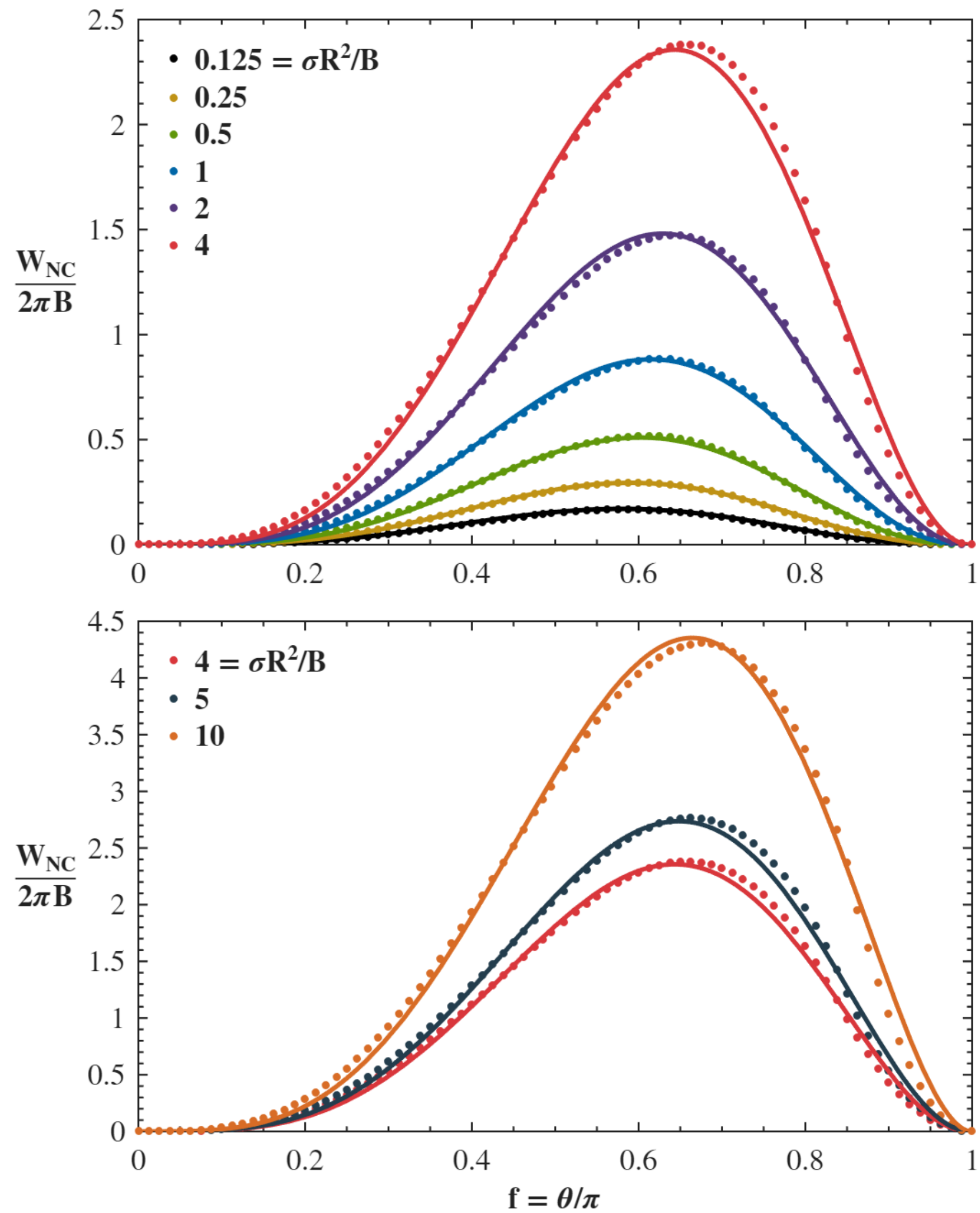

**Figure A1**: Dimensionless free energy of membrane deflection in the NC domain $W_{NC}/2\pi B$ versus wrapping fraction $f = \theta/\pi$ for varying $\bar{\sigma} = \sigma R^2/B$. The markers show the solution of Eq. (12) via numerical integration, while solid lines indicate the results of the fitting functions in Eqs. (A1)-(A5). Each fit employs eighty evenly spaced data points ($\Delta f = 0.0125$) while excluding $f = 0.5$ (where the numerical shooting solution does not converge).

Fig. A1 presents the dimensionless NC energy $W_{NC}/2\pi B$ vs the wrapping degree $f = \theta/\pi$ for a range of $\bar{\sigma} = 0.125, 0.25, 0.5, 1, 2, 4, 5,$ and $10$. The fitted profiles are bell-shaped, with the squared width given by

$$\Delta_f^2 = \frac{f^*(1-f^*)}{k-1} \tag{A6}$$

The circular markers in Fig. A1 represent the simulated energy values obtained by numerically integrating Eq. (12) for each $f$, with a total of 80 data points for each $\bar{\sigma}$ value. The solid lines show the analytical fits from Eqs. (A1)–(A5). For the catenoid case ($\bar{\sigma} = 0$), the NC energy is trivially zero throughout the wrapping process and is hence omitted from Figure A1. As $\bar{\sigma}$ increases, the overall NC energy rises substantially at every wrapping degree. Here, we mainly observe that the peak NC energy $W_{NC}^*$ occurs after the mid-wrap point: $f^* > 0.5$, *i.e.*, $\theta^* = \pi f^* > \pi/2$. Also, $f^*$ shifts progressively to higher values as $\bar{\sigma}$ increases.

Fig. A2-*left* shows $W_{NC}^*$ versus $\bar{\sigma}$, in a log-log plot. The power law in Eq. (A3) is inspired by the fit in this figure, which is executed following two scaling regimes based on the value of $\bar{\sigma}$. For $\bar{\sigma} \leq 1$, Fig. A2-*left* shows the red line fit, while for $\bar{\sigma} > 1$, the purple line follows the same equation but with different power-law coefficients. These coefficients, $a$ and $\alpha$, are reported in Table A1, obtained via least-squares fit, with an accuracy of $R^2 > 0.999$.

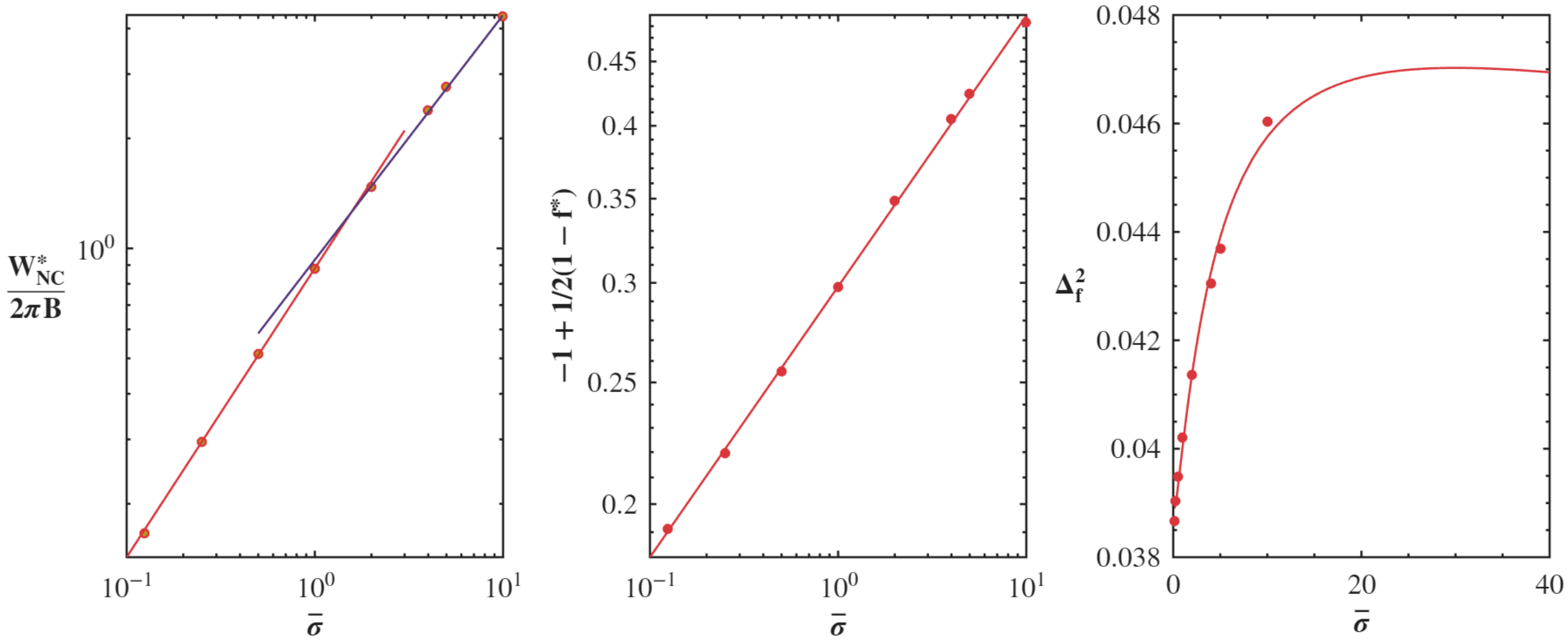

**Figure A2**: Fits of the tension-dependent parameters in the NC-energy approximation.

Fig. A2-*center* shows $f^*$ versus $\bar{\sigma}$, in a semi-log-x plot, where the fitted curve validates Eq. (A4). In this equation we assume that $f^* = 0.5$ at $\bar{\sigma} = 0$, while $f^* \approx 1$ for $\bar{\sigma} \to \infty$. The coefficients $b$ and $\beta$, from Eq. (A4), are reported in Table A1, and obtained via least-squares fit with $R^2 = 0.9994$.

Fig. A2-*right* shows $\Delta_f^2$ versus $\bar{\sigma}$, in a linear plot. Here we compare our results with the fitting function in Eq. (A6), with substitution from (A4) and (A5). The forms of Eqs. (A5) and (A6) are chosen to ensure that the bell width $\Delta_f$ decays to 0 as $\bar{\sigma} \to \infty$. The coefficients $c$, $d$, and $e$ are reported in Table A1, obtained via least-squares fit, with an accuracy of $R^2 = 0.9979$.

**Table A1**: Fitting coefficients of Eqs. (A3)–(A5).

| $a$ ($\bar{\sigma} \leq 1$) | $a$ ($\bar{\sigma} > 1$) | $\alpha$ ($\bar{\sigma} \leq 1$) | $\alpha$ ($\bar{\sigma} > 1$) | $b$ | $\beta$ | $c$ | $d$ | $e$ |
|---|---|---|---|---|---|---|---|---|
| 0.8824 | 0.9507 | 0.7910 | 0.6596 | 0.2980 | 0.2154 | 5.3032 | 7.9553 | 3.9374 |

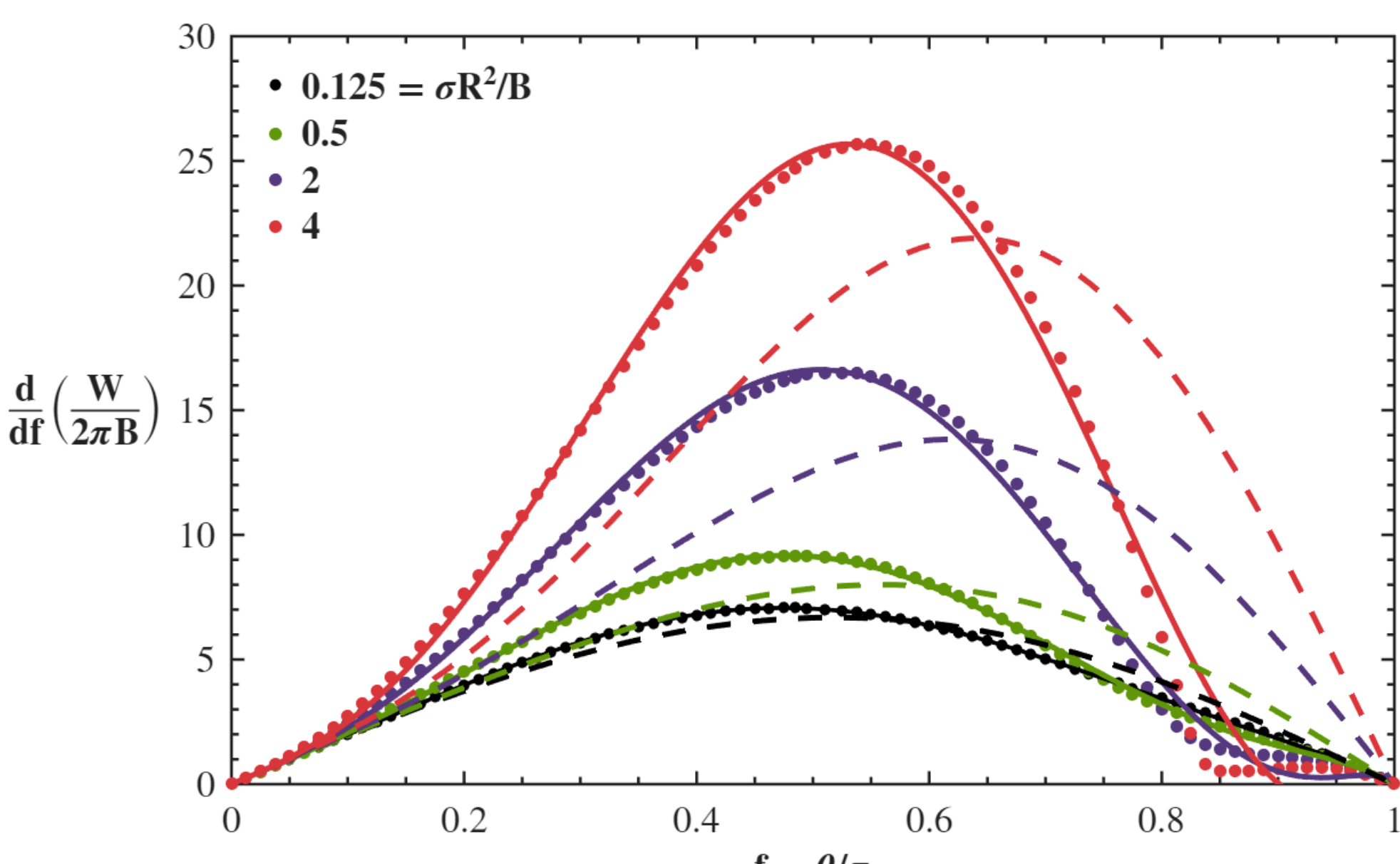

**Figure A3**: Rate of change of the dimensionless membrane free energy reported in Fig. 4, $d\bar{W}/df$, as a function of wrapping fraction $f$ (Eq. (25)) for different values of the dimensionless membrane tension $\bar{\sigma} = \sigma R^2/B$. This quantity represents the energetic cost per unit increase in wrapping fraction that must be overcome for wrapping to proceed spontaneously. Markers denote derivatives obtained from the numerical energy values, whereas the solid curves are obtained by differentiating the fitted NC-energy approximation presented in *Appendix A* and adding the analytical C-domain contribution. Dashed curves show the C-domain contribution alone and therefore neglect the NC-domain energy rate. The difference between the solid and dashed curves demonstrates that deformation of the NC membrane substantially modifies the local energetic requirement for wrapping, particularly at intermediate wrapping fractions.

In Fig. A3, we plot the rate of change of the total dimensionless membrane free energy, $\partial\bar{W}/\partial f$, with $\bar{W} = W/2\pi B$ and $W = W_C + W_{NC}$, as a function of wrapping fraction $f$ (Eq. (25)), for $\bar{\sigma} = 0.125, 0.5, 2,$ and $4$. The total energy rate is decomposed as $\partial\bar{W}/\partial f = \partial\bar{W}_C/\partial f + \partial\bar{W}_{NC}/\partial f$. The markers represent finite-difference derivatives of the numerical energy values, whereas the solid curves are obtained by differentiating the fitted NC-energy function from Eq. (A1) and adding the analytical C-domain contribution, from Eq. (17b). The dashed curves show $\partial\bar{W}_C/\partial f$ alone and therefore neglect the NC-domain contribution. The discrepancy between the solid and dashed curves quantifies the error introduced by ignoring the energy rate associated with deformation of the NC membrane. For wrapping to proceed spontaneously, the membrane-deflection energy rate must be overcome by the rate of adhesive-energy release at every wrapping fraction.

## Acknowledgments

This work was supported by the Natural Sciences and Engineering Research Council of Canada (NSERC) (RGPIN-2025-07085).